\documentclass[12pt]{article}
\usepackage{amsmath}
\usepackage{graphicx}
\usepackage{cite}
\usepackage{url}
\usepackage{xcolor}
\usepackage[colorlinks,citecolor=blue,urlcolor=blue,linkcolor=blue]{hyperref}

\newcommand\pubnumber{}
\newcommand\pubdate{\today}

\def\institute{Physics Department, Florida State University\\
Tallahassee, FL 32306-4350, USA}
\def\support{\footnote{Work is supported in part by US Department of Energy
under grant DE-SC010102 and by an intramural award from the FSU Provost’s
Office and the OPDA.}}

\def\Title#1{\begin{center} {\Large #1 } \end{center}}
\def\Author#1{\begin{center}{ \sc #1} \end{center}}
\def\Address#1{\begin{center}{ \it #1} \end{center}}

\newcommand\pubblock{\rightline{\begin{tabular}{l} \pubnumber\\
         \pubdate  \end{tabular}}}
\newenvironment{Abstract}{\begin{quotation}  }{\end{quotation}}
\newenvironment{Presented}{\begin{quotation} \begin{center} 
             PRESENTED AT\end{center}\bigskip 
      \begin{center}\begin{large}}{\end{large}\end{center} \end{quotation}}

\def\beq{\begin{equation}}
\def\eeq#1{\label{#1}\end{equation}}
\def\eeqn{\end{equation}}

\def\beqa{\begin{eqnarray}}
\def\eeqa#1{\label{#1}\end{eqnarray}}
\def\eeqan{\end{eqnarray}}

\let\bar=\overbar

\def\Dslash{\not{\hbox{\kern-4pt $D$}}}
\def\dslash{\not{\hbox{\kern-2pt $\del$}}}

\def\msb{{\bar{\ssstyle M \kern -1pt S}}}

\def\powhegbox{\textsc{Powheg-Box}}
\def\mg5{\textsc{MG5\_aMC@NLO}}
\def\fxfx{\textsc{FxFx}}

\begin{document}
\begin{titlepage}
\pubblock

\vfill
\Title{Theory advances for $t\bar{t}W$ multi-lepton predictions}
\vfill
\Author{ Manfred Kraus\support}
\Address{\institute}
\vfill
\begin{Abstract}
We report on recent theoretical advances in the description of the $pp\to
t\bar{t}W$ process. First, we discuss a comparison of many state-of-the-art
predictions for multi-lepton signatures including the leading QCD contributions
at $\mathcal{O}(\alpha_s^3\alpha^6)$ as well as subleading EW contributions at
$\mathcal{O}(\alpha_s\alpha^8)$. Furthermore, we briefly discuss recent
improvements using multi-jet merging techniques.
\end{Abstract}
\vfill
\begin{Presented}
$15^\mathrm{th}$ International Workshop on Top Quark Physics\\
Durham, UK, 4--9 September, 2022
\end{Presented}
\vfill
\end{titlepage}
\def\thefootnote{\fnsymbol{footnote}}
\setcounter{footnote}{0}
%
\section{Introduction}
The production of $t\bar{t}W$ final states is one of the rarest processes at
the LHC.  At the same time, the various decay channels constitute one of the
most important backgrounds in $t\bar{t}H$ measurements as well as in searches
for $t\bar{t}t\bar{t}$ signatures.  In recent years, the production of
top-quark pairs along with a $W^\pm$ gauge boson has received much attention as
tensions between data and theoretical predictions have been
reported~\cite{ATLAS-CONF-2019-045,CMS-PAS-HIG-19-008}.

While the first calculation for on-shell $t\bar{t}W$ appeared a decade
ago~\cite{Campbell:2012dh}, more refined predictions have emerged recently.
Besides electroweak effects~\cite{Frixione:2015zaa,Frederix:2018nkq,
Dror:2015nkp,Frederix:2017wme} also the resummation of soft gluons has been
addressed in Refs.~\cite{Li:2014ula,Broggio:2016zgg,Kulesza:2018tqz,
Broggio:2019ewu}. The $pp\to t\bar{t}W$ process has been also matched to parton
showers using both, the MC@NLO~\cite{Maltoni:2014zpa,Maltoni:2015ena,
Frederix:2020jzp} as well as the POWHEG method~\cite{Garzelli:2012bn,
FebresCordero:2021kcc}. Additionally, also the impact of multi-jet merging has
been studied in Refs.~\cite{vonBuddenbrock:2020ter,Frederix:2021agh}. Beyond the 
stable top-quark approximation the size of off-shell effects and non-resonant
contributions have been investigated. First for the leading QCD corrections
\cite{Bevilacqua:2020pzy,Denner:2020hgg,Bevilacqua:2020srb} and afterwards also
for the electroweak contributions~\cite{Denner:2021hqi,Bevilacqua:2021tzp}.

Here, we will briefly summarize the recent progress made in the field at hand
of two examples from Refs.~\cite{Bevilacqua:2021tzp,Frederix:2021agh}.
\section{Phenomenological Results}
First we discuss the recent comparison of many state-of-the-art Standard Model
predictions. For details of the comparison we refer to
Ref.~\cite{Bevilacqua:2021tzp}.
\begin{table}[h]
\centering
\begin{tabular}{c|cc}
 $\sigma^{\textrm{NLO}}_{\textrm{QCD}}$ & $t\bar{t}W$ QCD [fb] & $t\bar{t}W$ EW [fb] \\
\hline
full off-shell     & $1.58^{+3\%}_{-6\%}$   & $0.206^{+22\%}_{-17\%}$ \\
full NWA           & $1.57^{+3\%}_{-6\%}$   & $0.190^{+22\%}_{-16\%}$ \\
NWA with LO decays & $1.66^{+10\%}_{-10\%}$ & $0.162^{+22\%}_{-16\%}$ \\
\hline
\powhegbox{}       & $1.40^{+11\%}_{-11\%}$ & $0.133^{+21\%}_{-16\%}$ \\
\mg5{}             & $1.40^{+11\%}_{-11\%}$ & $0.136^{+21\%}_{-16\%}$
\end{tabular}
\caption{Fiducial cross sections for $t\bar{t}W$ at $\mathcal{O}(\alpha_s^3\alpha^6)$ (QCD) and for
$\mathcal{O}(\alpha_s\alpha^8)$ (EW).}
\label{tab:xsec}
\end{table}
In Tab.~\ref{tab:xsec} the fiducial cross sections are listed.  First, we note
that the parton-shower based predictions agree well with each other, while they
predict consistently lower cross sections as comparable fixed-order approaches.
This difference originates from multiple emissions in the top-quark decays that
effects the acceptance efficiency in the fiducial volume. Furthermore, we
observe that for $t\bar{t}W$ QCD the theoretical uncertainty of parton shower
event generators are larger than in the case of the full off-shell or
narrow-width approximation (NWA) predictions due to missing higher-order
corrections in the decays. For $t\bar{t}W$ EW predictions all the effects are
much more pronounced because even the full NWA performs badly and off-shell
effects are already of the order of $9\%$ at the level of inclusive cross
sections. Even though the theoretical uncertainties are much larger for
$t\bar{t}W$ EW and modeling issues are prominent they only play a minor role in
the end as the EW production mode only contributes $13\%$ of the leading QCD
cross section.

\begin{figure}[h!]
 \centering
 \includegraphics[width=0.48\textwidth]{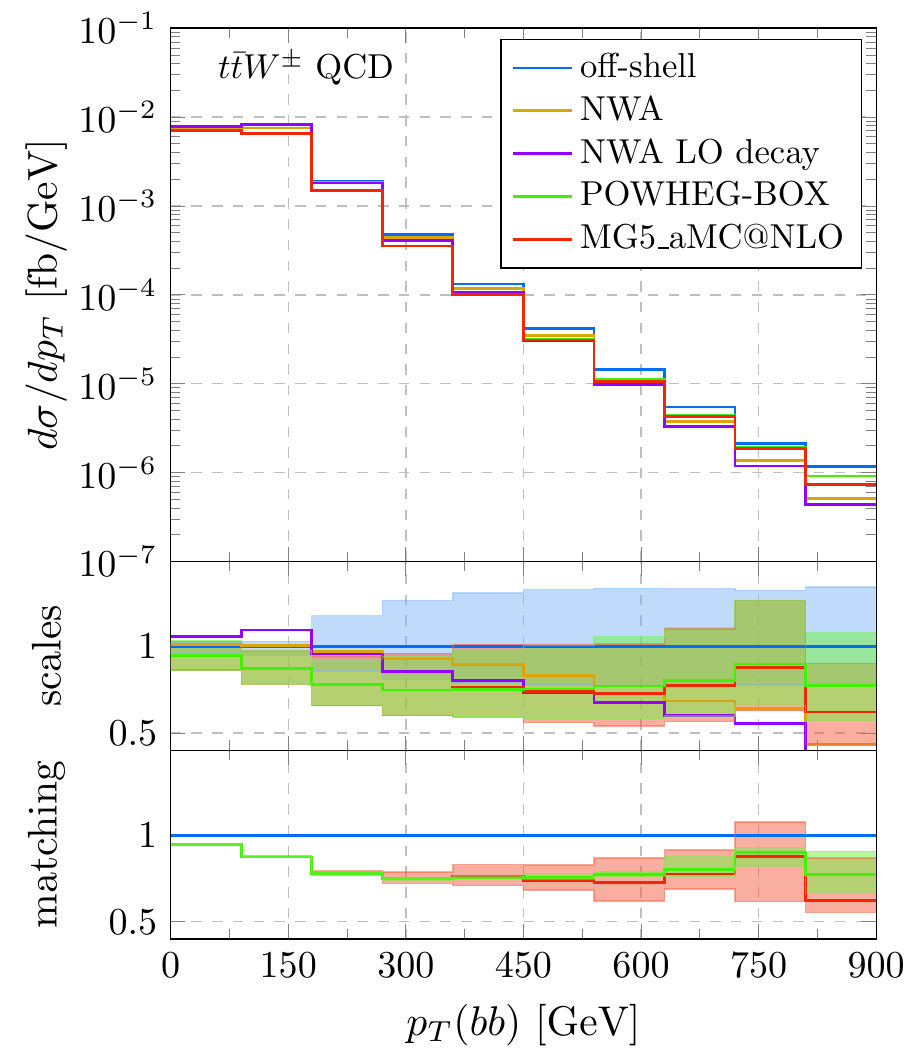}
 \includegraphics[width=0.48\textwidth]{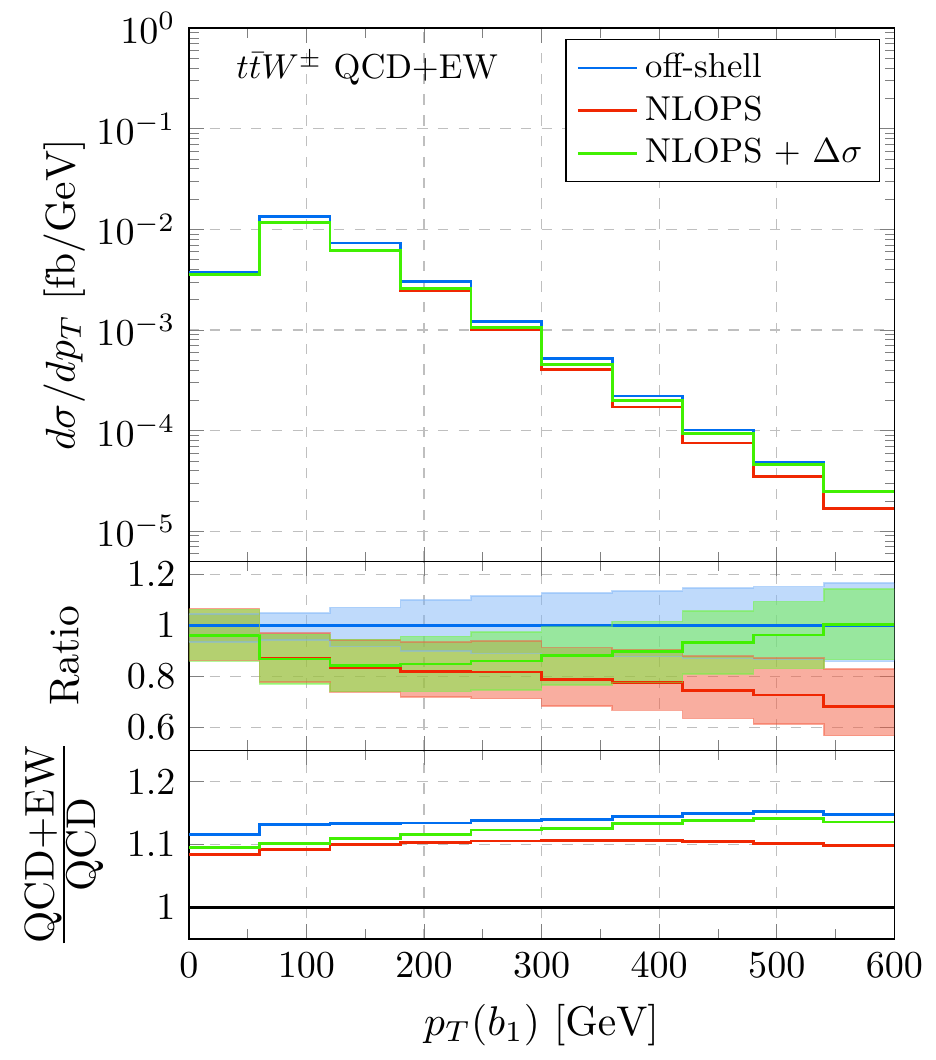}
 \caption{Differential cross sections for various Standard model predictions.
Plots taken from Ref.~\cite{Bevilacqua:2021tzp}.}
 \label{fig:diff_1}
\end{figure}
On the left side of Fig.~\ref{fig:diff_1} the transverse momentum of the
leading $b\bar{b}$ pair is shown for various theoretical predictions. While
even at the differential level the parton shower based results are very
consistent with each other, non of the predictions resembles the shape of the
full off-shell calculation. While the full NWA still agrees in the bulk of the
distribution, all predictions deviate strongly in the tail which is dominated
by single-resonant contributions that are only included in the full off-shell
prediction.
The right plot of Fig.~\ref{fig:diff_1} depicts improved predictions for the
transverse momentum of the leading $b$ jet. Here, parton shower matched
calculations are supplemented with single and non-resonant contributions via
\begin{equation}
 \frac{d\sigma^{\textrm{th}}}{dX} = \frac{d\sigma^{\textrm{NLOPS}}}{dX} + 
 \frac{d\Delta\sigma_{\textrm{off-shell}}}{dX}\;, \qquad
 \frac{d\Delta\sigma_{\textrm{off-shell}}}{dX} = 
 \frac{d\sigma^{\textrm{NLO}}_{\textrm{off-shell}}}{dX} - 
 \frac{d\sigma^{\textrm{NLO}}_{\textrm{NWA}}}{dX}\;
\end{equation}
and denoted with NLOPS + $\Delta\sigma$. First of all, we note that the parton
shower matched predictions (NLOPS) differ with respect to the full off-shell
result up to $35\%$ in the tail of the distribution. However, once single and
non-resonant contributions are taken into account via the differential
corrections $d\Delta\sigma_{\textrm{off-shell}}/dX$ the improved predictions
agree well with the full off-shell calculation. The bottom panel also shows
that the electroweak contribution receives a sizable correction in the tail of
the distribution.

\section{Multi-jet Merging}
An orthogonal approach to increase the accuracy of theoretical predictions is
multi-jet merging, where multiple parton-shower matched calculations for $pp\to
t\bar{t}W$, $pp\to t\bar{t}Wj$ and $pp\to t\bar{t}Wjj$ are combined into a
single prediction. Here we want to briefly mention the improved
\fxfx{}~\cite{Frederix:2012ps} merging for $pp\to
t\bar{t}W$~\cite{Frederix:2021agh}.
\begin{figure}[h!]
 \centering
 \includegraphics[width=0.4\textwidth]{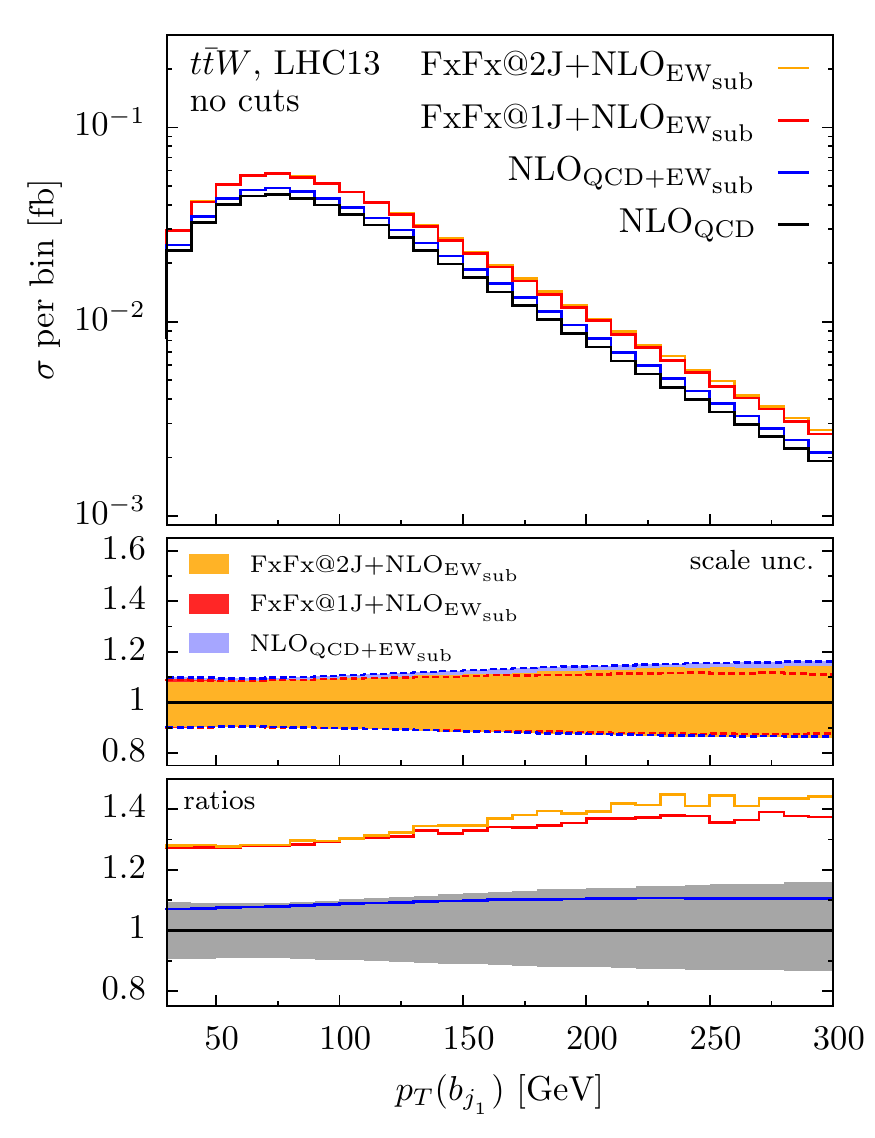}
 \includegraphics[width=0.4\textwidth]{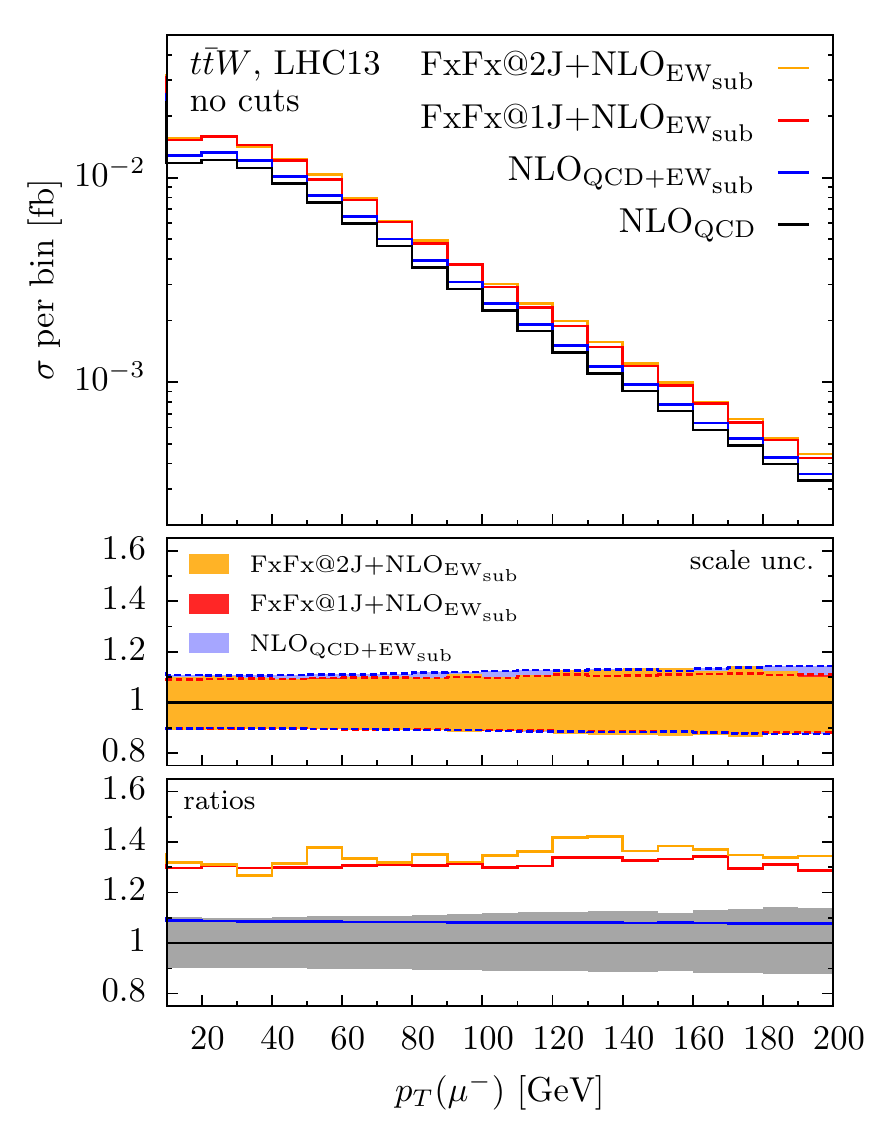}
 \caption{Differential distribution for the transverse momentum of the leading
$b$ jet and the muon. Plots taken from Ref.~\cite{Frederix:2021agh}.}
 \label{fig:diff_2}
\end{figure}
The main advancement of Ref.~\cite{Frederix:2021agh} relies on the
identification of so-called QCD and electroweak jets. Here, QCD jets originate
from QCD splittings such as $q\to qg$, while EW jets are formed by EW
splittings such as $q\to W q^\prime$.  The latter splittings are finite for
small transverse momenta and therefore a resummation via parton showers is not
necessary. 

Based on this improvement, we show in Fig.~\ref{fig:diff_2} improved \fxfx{}
predictions for $t\bar{t}W$. In particular the transverse momentum of the
leading $b$ jet and the muon is shown. It is evident from the bottom panel that
the inclusion of higher jet multiplicities introduces large corrections between
$30-40\%$. Therefore, the inclusion of additional hard jet radiation is
necessary to obtain an accurate description of $t\bar{t}W$ in fiducial volumes.

\section{Conclusions}
We have summarized recent progress for theoretical predictions for the $pp\to
t\bar{t}W$ process. The main conclusions that can be drawn are as follows:
\begin{itemize}
 \item The theoretical uncertainties of currently available parton-shower
matched predictions might be reduced by including QCD corrections to top-quark
decays.

 \item Multiple radiation in the top-quark decays has a strong impact in
fiducial volumes.

 \item Contributions from additional hard jets are sizable and should be
included.

 \item Tails of dimensionful distributions are often dominated by single and
non-resonant contributions.
\end{itemize}
All these observations essentially point to the fact that the $pp\to t\bar{t}W$
process needs to be computed at the NNLO accuracy including top-quark decays
via the narrow-width approximation. Furthermore, a full off-shell calculation
for $t\bar{t}W$ matched to parton showers is highly desirable to include
further corrections. 

\bibliography{main}
\bibliographystyle{JHEP}
 
\end{document}